\documentclass{article}
\usepackage{ismir,amsmath,cite}
\usepackage{graphicx}
\usepackage{color}
\usepackage{mathrsfs}
\usepackage[english]{babel}
\usepackage{caption}
\usepackage{subfig, color}
\usepackage{microtype}
\usepackage{IEEEtrantools}
\usepackage{url}
\usepackage{amsfonts}
\usepackage{xspace}
\newcommand*{\eg}{e.g.\@\xspace}
\newcommand*{\ie}{i.e.\@\xspace}
\newcommand*{\etal}{et al.\@\xspace}
\newcommand*{\s}[1]{\small{($\pm$ #1)}}
\newcommand*{\g}[1]{\small{\textbf{(}$\boldsymbol{\pm}$ \textbf{#1}\textbf{)}}}

\usepackage{lipsum}

\newcommand\blfootnote[1]{%
  \begingroup
  \renewcommand\thefootnote{}\footnote{#1}%
  \addtocounter{footnote}{-1}%
  \endgroup
}

\title{Deep convolutional networks on the pitch spiral for musical instrument recognition}

\oneauthor
{Vincent Lostanlen and Carmine-Emanuele Cella}
{\'{E}cole normale sup\'{e}rieure,
PSL Research University,
CNRS,
Paris, France}

\begin{document}
\maketitle
\begin{abstract}
Musical performance combines a wide range of pitches, nuances,
and expressive techniques.
Audio-based classification of musical instruments thus requires to
build signal representations that are invariant to such transformations.
This article investigates the construction
of learned convolutional architectures for instrument recognition,
given a limited amount of annotated training data.
In this context, we benchmark three different weight sharing
strategies for deep convolutional networks in the
time-frequency domain: temporal kernels;
time-frequency kernels;
and a linear combination of time-frequency
kernels which are one octave apart, akin to a Shepard pitch spiral.
We provide an acoustical interpretation of these strategies
within the source-filter framework of quasi-harmonic sounds with
a fixed spectral envelope, which are archetypal of musical notes.
The best classification accuracy is obtained by hybridizing all three
convolutional layers into a single deep learning architecture.
\blfootnote{This work is supported by the ERC InvariantClass grant 320959.
The source code to reproduce figures and experiments is freely available at
\url{www.github.com/lostanlen/ismir2016}.}

\end{abstract}

\section{Introduction}\label{sec:introduction}
Among the cognitive attributes of musical tones, pitch is distinguished
by a combination of three properties.
First, it is relative: ordering pitches from low to high gives rise to
intervals and melodic patterns.
Secondly, it is intensive: multiple pitches heard simultaneously produce
a chord, not a single unified tone -- contrary to loudness, which adds
up with the number of sources.
Thirdly, it does not depend on instrumentation: this makes possible
the transcription of polyphonic music under a single symbolic system
\cite{deCheveigne2005}.

Tuning auditory filters to a perceptual scale of pitches provides a
time-frequency representation of music signals that satisfies the first two of these properties.
It is thus a starting point for a wide range of MIR applications,
which can be separated in two categories: pitch-\emph{relative}
(\eg chord estimation \cite{Humphrey2012tonnetz})
and pitch-\emph{invariant} (\eg instrument recognition \cite{Eronen2000}).
Both aim at disentangling pitch from timbral content as independent
factors of variability, a goal that is made possible by the third aforementioned property.
This is pursued by extracting mid-level features on top of the spectrogram,
be them engineered or learned from training data.
Both approaches have their limitations: a "bag-of-features" lacks flexibility
to represent fine-grain class boundaries, whereas a purely learned pipeline
often leads to uninterpretable overfitting, especially in MIR where the quantity
of thoroughly annotated data is relatively small.

In this article, we strive to integrate domain-specific knowledge about musical
pitch into a deep learning framework, in an effort towards bridging the gap
between feature engineering and feature learning.

Section 2 reviews the related work on feature learning for signal-based music
classification.
Section 3 demonstrates that pitch is the major factor of variability among musical
notes of a given instrument, if described by their mel-frequency cepstra.
Section 4 presents a typical deep learning architecture for spectrogram-based
classification, consisting of two convolutional layers in the time-frequency
domain and one densely connected layer.
Section 5 introduces alternative convolutional architectures for learning
mid-level features, along time and along a Shepard pitch spiral, as well as
aggregation of multiple models in the deepest layers.
Sections 6 discusses the effectiveness of the presented systems on a challenging
dataset for musical instrument recognition.

\section{Related work}
Spurred by the growth of annotated datasets and the democratization of
high-performance computing, feature learning has enjoyed a renewed interest
in recent years within the MIR community, both in supervised and unsupervised
settings.
Whereas unsupervised learning (\eg $k$-means \cite{Stowell2014}, Gaussian
mixtures \cite{Joder2009}) is employed to fit the distribution of the data with
few parameters of relatively low abstraction
and high dimensionality, state-of-the-art supervised learning consists of a deep
composition of multiple nonlinear transformations, jointly optimized
to predict class labels, and whose behaviour tend to gain in abstraction as depth
increases \cite{vandenOord2013}.

As compared to other deep learning techniques for audio processing,
convolutional networks happen to strike the balance between
learning capacity and robustness.
The convolutional structure of learned transformations is derived from
the assumption that the input signal, be it a one-dimensional waveform
or a two-dimensional spectrogram, is stationary --- which means that
content is independent from location.
Moreover, the most informative dependencies between signal coefficients
are assumed to be concentrated to temporal or spectrotemporal neighborhoods.
Under such hypotheses, linear transformations can be learned efficiently
by limiting their support to a small kernel which is convolved over
the whole input.
This method, known as weight sharing, decreases the number of parameters
of each feature map while increasing the amount of data on which kernels are
trained.

By design, convolutional networks seem well adapted to instrument
recognition, as this task does not require a precise timing of the activation function,
and is thus essentially a challenge of temporal integration \cite{Eronen2000, Joder2009}.
Furthermore, it benefits from an unequivocal ground truth, and may be simplified
to a single-label classification problem by extracting individual stems from a
multitrack dataset \cite{Bittner2014}.
As such, it is often used a test bed for the development of new algorithms
\cite{McFee2015-muda, Li2015},
as well as in computational studies in music cognition \cite{Newton2012, Patil2012}.

Some other applications of deep convolutional networks include onset
detection \cite{Schluter2014}, transcription \cite{Sigtia2015},
chord recognition \cite{Humphrey2012tonnetz},
genre classification \cite{Choi2015},
downbeat tracking \cite{Durand2016},
boundary detection \cite{Ullrich2014}, and
recommendation \cite{vandenOord2013}.

Interestingly, many research teams in MIR have converged to employ the same
architecture, consisting of two convolutional layers and two densely connected layers
\cite{Dieleman2014, Humphrey2012tonnetz,
Kereliuk2015, Li2015, McFee2015-muda, Schluter2014, Ullrich2014}, and this
article makes no exception.
However, there is no clear consensus regarding the
weight sharing strategies that should be applied to musical audio streams:
convolutions in time or in time-frequency coexist in the recent literature.
A promising paradigm \cite{Dieleman2011, Durand2016}, at the interaction
between feature engineering and feature learning, is to extract
temporal or spectrotemporal descriptors of various low-level modalities,
train specific convolutional layers on each modality
to learn mid-level features, and hybridize information at the top level.
Recognizing that this idea has been successfully applied to large-scale artist
recognition \cite{Dieleman2011} as well as downbeat tracking
\cite{Durand2016}, we aim to proceed in a comparable way for instrument
recognition.

\section{How invariant is the Mel-frequency cepstrum ?}
The mel scale is a quasi-logarithmic function of acoustic frequency designed such that
perceptually similar pitch intervals appear equal in width over the full hearing range.
This section shows that engineering transposition-invariant features from the mel
scale does not suffice to build pitch invariants for complex sounds, thus motivating
further inquiry.

The time-frequency domain produced by a constant-Q filter bank tuned to the mel
scale is covariant with respect to pitch transposition of pure tones.
As a result, a chromatic scale played at constant speed would draw parallel,
diagonal lines, each of them corresponding to a different partial wave.
However, the physics of musical instruments constrain these partial waves to bear
a negligible energy if their frequencies are beyond the range of acoustic resonance.

\begin{figure}[t]
    \begin{center}
        \setlength{\unitlength}{1cm}
        \begin{picture}(8.2,3)
        \put(0,-0.5){\includegraphics[width=8cm]{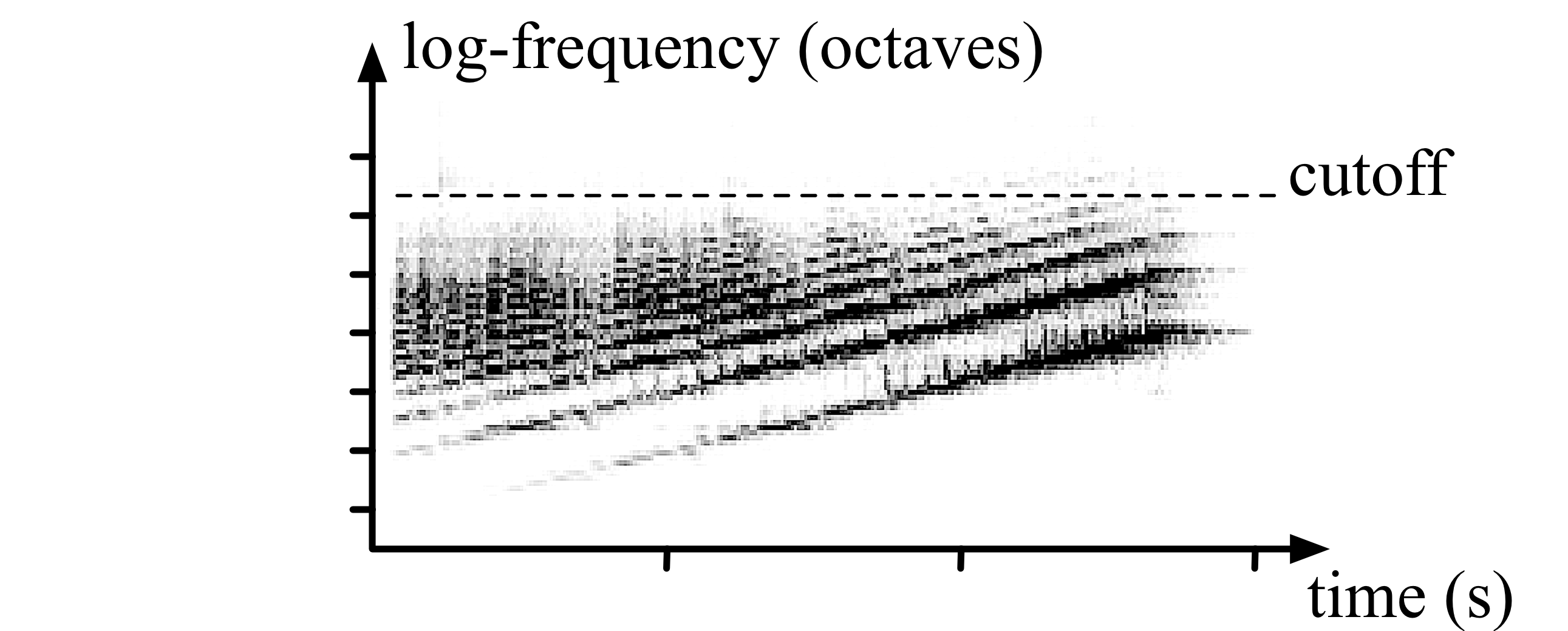}}
        \end{picture}
    \end{center}
    \protect\caption{
    Constant-Q spectrogram of a chromatic scale played by a tuba.
    Although the harmonic partials shift progressively, the spectral envelope remains unchanged,
    as revealed by the presence of a fixed cutoff frequency.
    See text for details.
\label{fig:chromatic-scale}
}
\end{figure}

As shown on Figure \ref{fig:chromatic-scale}, the constant-Q spectrogram of a
tuba chromatic scale exhibits a fixed,
cutoff frequency at about $2.5\,\mathrm{kHz}$, which
delineates the support of its spectral envelope.
This elementary observation implies that realistic pitch changes cannot be modeled
by translating a rigid spectral template along the log-frequency axis.
The same property is verified for a wide class of instruments, especially brass and
woodwinds.
As a consequence, the construction of powerful invariants to musical pitch is not
amenable to delocalized operations on the mel-frequency spectrum, such as a
discrete cosine transform (DCT) which leads to the mel-frequency cepstral
coefficients (MFCC), often used in audio classification \cite{Eronen2000, Joder2009}.

To validate the above claim, we have extracted the MFCC
of 1116 individual notes from the RWC dataset \cite{Goto2003},
as played by 6 instruments, with
32 pitches, 3 nuances,
and 2 interprets and manufacturers.
When more than 32 pitches were available (\eg piano), we selected
a contiguous subset of 32 pitches in the middle register.
Following a well-established rule \cite{Eronen2000, Joder2009},
the MFCC were defined the 12 lowest nonzero "quefrencies" among the
DCT coefficients extracted from a filter bank of 40 mel-frequency bands.
We then have computed the distribution of squared Euclidean distances
between musical notes in the 12-dimensional space of MFCC features.

\begin{figure}[t]
    \begin{center}
        \setlength{\unitlength}{1cm}
        \begin{picture}(8.5,8.7)
        \put(0.1,0){\includegraphics[width=8cm]{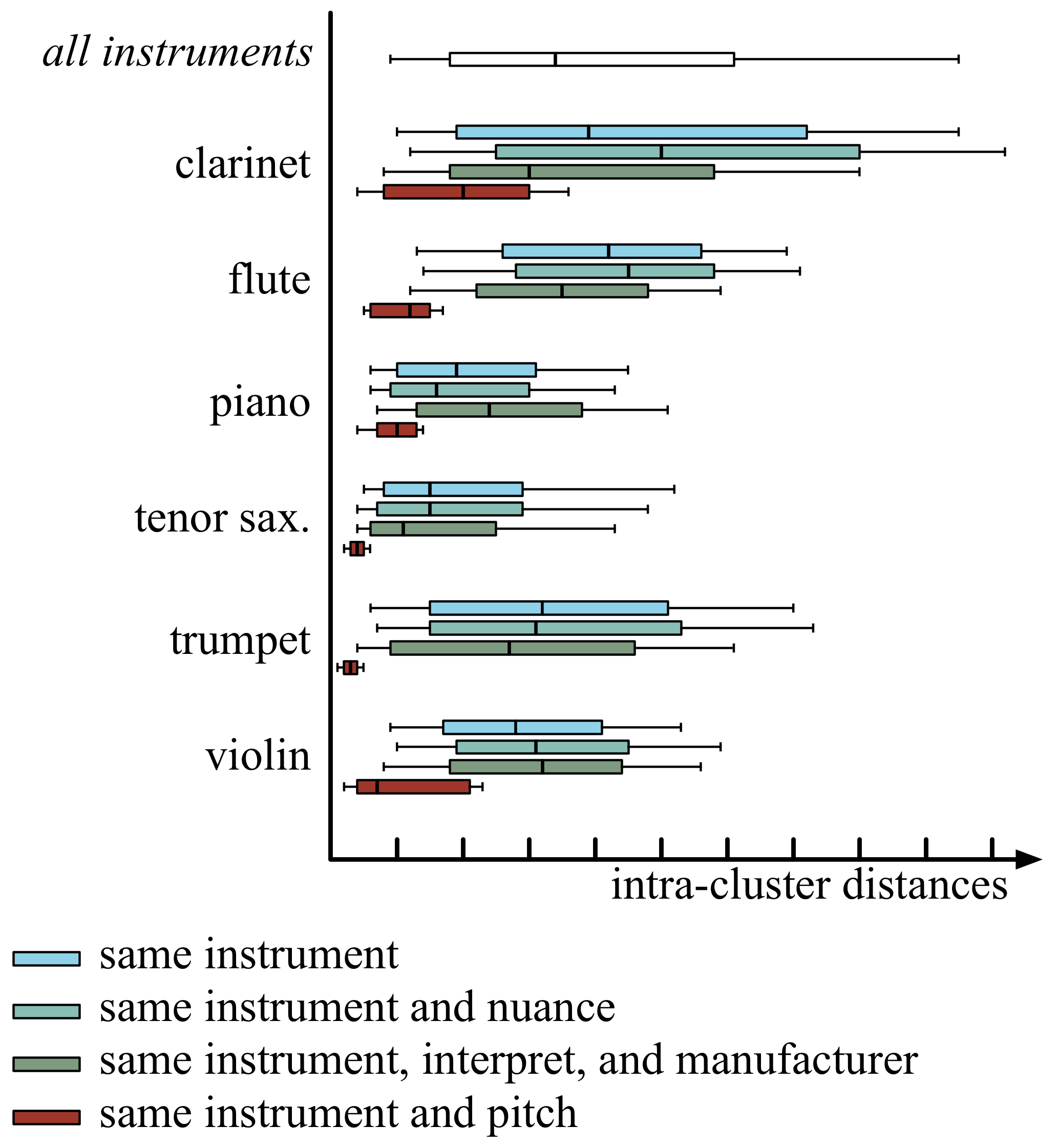}}
        \end{picture}
    \end{center}
    \protect\caption{
Distributions of squared Euclidean distances among various MFCC clusters in the RWC dataset.
Whisker ends denote lower and upper deciles. See text for details.
\label{fig:mfcc-variances}
}
\end{figure}

Figure \ref{fig:mfcc-variances} summarizes our results.
We found that restricting the cluster to one nuance, one interpret, or one manufacturer
hardly reduces intra-class distances.
This suggests that MFCC are fairly successful in building invariant representations
to such factors of variability.
In contrast, the cluster corresponding to each instrument is shrinked if
decomposed into a mixture of same-pitch clusters, sometimes by an order of
magnitude.
In other words, most of the variance in an instrument cluster of mel-frequency
cepstra is due to pitch transposition.

Keeping less than 12 coefficients certainly improves invariance, yet at the cost of
inter-class discriminability, and vice versa.
This experiment shows that the mel-frequency cepstrum is perfectible in terms
of invariance-discriminability tradeoff, and that there remains a lot to be gained by
feature learning in this area.
\section{Deep convolutional networks}
A deep learning system for classification is built by stacking multiple layers of weakly nonlinear
transformations, whose parameters are optimized such that the top-level layer fits a training
set of labeled examples.
This section introduces a typical deep learning architecture for audio classification and describes
the functioning of each layer.

\begin{figure*}[t]
    \begin{center}
        \setlength{\unitlength}{1cm}
        \begin{picture}(17,5)
        \put(0,0){\includegraphics[width=17cm]{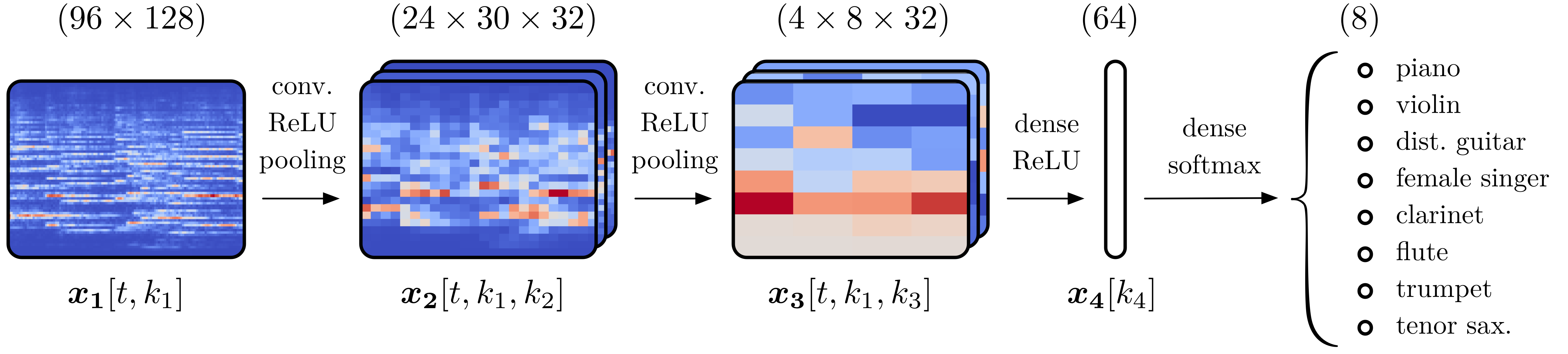}}
        \end{picture}
    \end{center}
    \protect\caption{
A two-dimensional deep convolutional network trained on constant-Q spectrograms. See text for details.
\label{fig:instrument-distribution}
}
\end{figure*}

Each layer in a convolutional network typically consists in the composition of three operations:
two-dimensional convolutions, application of a pointwise nonlinearity, and local pooling.
The deep feed-forward network made of two convolutional layers and two densely connected
layers, on which our experiment are conducted,
has become a \emph{de facto} standard in the MIR community
\cite{Dieleman2014, Humphrey2012tonnetz,
Kereliuk2015, Li2015, McFee2015-muda, Schluter2014, Ullrich2014}.
This ubiquity in the literature suggests that a four-layer network with two convolutional
layers is well adapted to supervised audio classification problems of moderate size.

The input of our system is a constant-Q spectrogram, which is very comparable to a
mel-frequency spectrogram.
We used the implementation from the librosa package \cite{McFee2015-librosa} with $Q=12$
filters per octave, center frequencies ranging from $\mathrm{A_1}$ ($55\,\mathrm{Hz}$)
to $\mathrm{A_9}$ ($14\,\mathrm{kHz}$), and a hop size of $23\,\mathrm{ms}$.
Furthermore, we applied nonlinear perceptual weighting of loudness in order to reduce the
dynamic range between the fundamental partial and its upper harmonics.
A $3$-second sound excerpt $\boldsymbol{x}[t]$ is represented by a time-frequency matrix
$\boldsymbol{x_1}[t,k_1]$ of width $T=128$ samples and height $K_1=96$ frequency bands.

A convolutional operator is defined as a family $\boldsymbol{W_2}[\tau,\kappa_1,k_2]$ of
$K_2$ two-dimensional filters, whose impulse repsonses are all constrained to have width
$\Delta t$ and height $\Delta k_1$. Element-wise biases $\boldsymbol{b_2}[k_2]$ are
added to the convolutions, resulting in the three-way tensor
\begin{IEEEeqnarray}{rCl}
\IEEEeqnarraymulticol{3}{l}{
\boldsymbol{y_2}[t,k_1,k_2]} \nonumber \\
& = & \boldsymbol{b_2}[k_2] +
\boldsymbol{W_2}[t, k_1, k_2] \overset{t,k_1}{\ast} \boldsymbol{x_1}[t, k_1]
\nonumber \\
& = &
\boldsymbol{b_2}[k_2] +
\sum_{\substack{
0 \leq \tau < \Delta t \\
0 \leq \kappa_1 < \Delta k_1}}
\! \! \! \! \!
\boldsymbol{W_2}[\tau, \kappa_1, k_2]
\boldsymbol{x_1}[t-\tau, k_1-\kappa_1].
\IEEEeqnarraynumspace
\label{eq:convolution2d}
\end{IEEEeqnarray}
The pointwise nonlinearity we have chosen is the rectified linear unit (ReLU),
with a rectifying slope of $\alpha=0.3$ for negative inputs.
\begin{IEEEeqnarray}{l}
\boldsymbol{y_{2}^{+}}[t,k_{1},k_{2}]=\left\{ \! \! \! \begin{array}{r}
\alpha\boldsymbol{y_{2}}[t,k_{1},k_{2}] \;\;\; \mbox{if} \;\; \boldsymbol{y_{2}}[t,k_{1},k_{2}]<0\\
\boldsymbol{y_{2}}[t,k_{1},k_{2}] \;\;\; \mbox{if} \;\; \boldsymbol{y_{2}}[t,k_{1},k_{2}]\geq0
\end{array}\right. \!
\IEEEeqnarraynumspace
\label{eq:relu}
 \end{IEEEeqnarray}
The pooling step consists in retaining the maximal activation among neighboring units in the
time-frequency domain $(t, k_1)$ over non-overlapping rectangles of width $\Delta t$ and
height $\Delta k_1$.
\begin{equation}
\boldsymbol{x_2}[t,k_1,k_2] = \! \!
\max_{
\substack{
0 \leq \tau < \Delta t \\
0 \leq \kappa_1 < \Delta k_1}
} \! \!
\left\{
\boldsymbol{y_{2}^{+}}[t - \tau, k_1 - \kappa_1, k_2]
\right\}
\label{eq:pooling}
\end{equation}
The hidden units in $\boldsymbol{x_2}$ are in turn fed to a second layer of convolutions,
ReLU, and pooling.
Observe that the corresponding convolutional operator
$\boldsymbol{W_3}[\tau, \kappa_1, k_2, k_3]$ performs a linear combination of time-frequency
feature maps in $\boldsymbol{x_2}$ along the variable $k_2$.
\begin{IEEEeqnarray}{rCl}
\IEEEeqnarraymulticol{3}{l}{
\boldsymbol{y_3}[t,k_1,k_3]} \nonumber \\
& = &
\sum_{k_2}
\boldsymbol{b_3}[k_2, k_3]
+ \boldsymbol{W_3}[t, k_1, k_2, k_3]
\overset{t,k_1}{\ast}
\boldsymbol{x_2}[t,k_1,k_2].
\IEEEeqnarraynumspace
\end{IEEEeqnarray}
Tensors $\boldsymbol{y_3^{+}}$ and $\boldsymbol{x_3}$ are derived from $\boldsymbol{y_3}$
by ReLU and pooling, with formulae similar to Eqs. (\ref{eq:relu}) and (\ref{eq:pooling}).
The third layer consists of the linear projection of $\boldsymbol{x_3}$, viewed as a vector of
the flattened index $(t, k_1, k_3)$, over $K_4$ units:
\begin{IEEEeqnarray}{rCl}
\boldsymbol{y_4}[k_4] =
\boldsymbol{b_4}[k_4] +
\sum_{t,k_1,k_3}
\boldsymbol{W_4}[t, k_1, k_3, k_4]
\boldsymbol{x_3}[t, k_1, k_3]
\label{eq:densely-connected-layer}
\IEEEeqnarraynumspace
\end{IEEEeqnarray}
We apply a ReLU to $\boldsymbol{y_4}$, yielding
$\boldsymbol{x_4}[k_4] = \boldsymbol{y_4^{+}}[k_4]$.
Finally, we project $\boldsymbol{x_4}$, onto a layer of output units $\boldsymbol{y_5}$ that
should represent instrument activations:
\begin{equation}
\boldsymbol{y_5}[k_5] = \sum_{k_4} \boldsymbol{W_5}[k_4, k_5] \boldsymbol{x_4}[k_4].
\end{equation}
The final transformation is a softmax nonlinearity, which ensures that output coefficients are
non-negative and sum to one, hence can be fit to a probability distribution:
\begin{equation}
\boldsymbol{x_5}[k_5] =
\frac{\exp \boldsymbol{y_5}[k_5]}
{  \sum_{\kappa_5} \exp \boldsymbol{y_5}[\kappa_5] }.
\end{equation}
Given a training set of spectrogram-instrument pairs $(\boldsymbol{x_1}, k)$,
all weigths in the network are iteratively updated to minimize the stochastic cross-entropy loss
$\mathscr{L}(\boldsymbol{x_5}, k) = - \log \boldsymbol{x_5}[k]$
over shuffled mini-batches of size $32$ with uniform class distribution.
The pairs $(\boldsymbol{x_1}, k)$ are extracted on the fly by selecting non-silent
regions at random within a dataset of single-instrument audio recordings.
Each $3$-second spectrogram $\boldsymbol{x_1}[t, k_1]$ within a batch is
globally normalized such that the whole batch has zero mean and unit variance.
At training time, a random dropout of 50\% is applied to the activations of
$\boldsymbol{x_3}$ and $\boldsymbol{x_4}$.
The learning rate policy for each scalar weight in the network is Adam \cite{Kingma2015},
a state-of-the-art online optimizer for gradient-based learning.
Mini-batch training is stopped after the average training loss stopped
decreasing over one full epoch of size $8192$.
The architecture is built using the Keras library \cite{Chollet2015}
and trained on a graphics processing unit within minutes.

\section{Improved weight sharing strategies}
Although a dataset of music signals is unquestionably stationary over the time
dimension -- at least at the scale of a few seconds -- it cannot be taken for granted
that all frequency bands of a constant-Q spectrogram would have the same
local statistics \cite{Humphrey2013}.
In this section, we introduce two alternative architectures to address the
nonstationarity of music on the log-frequency axis,
while still leveraging the efficiency of convolutional representations.

Many are the objections to the stationarity assumption among local neighborhoods
in mel frequency.
Notably enough, one of the most compelling is derived from the classical source-filter
model of sound production.
The filter, which carries the overall spectral envelope, is affected by intensity and
playing style, but not by pitch.
Conversely, the source, which consists of a pseudo-periodic wave, is transposed
in frequency under the action of pitch.
In order to extract the discriminative information present in both terms, it is first
necessary to disentangle the contributions of source and filter
in the constant-Q spectrogram.
Yet, this can only be achieved by exploiting long-range correlations in frequency,
such as harmonic and formantic structures.
Besides, the harmonic comb created by the Fourier series of the source makes an
irregular pattern on the log-frequency axis which is hard to characterize by local
statistics.

\subsection{One-dimensional convolutions at high frequencies}
Facing nonstationary constant-Q spectra,
the most conservative workaround is to increase the height $\Delta \kappa_1$ of each
convolutional kernel up to the total number of bins $K_1$ in the spectrogram.
As a result, $\boldsymbol{W_1}$ and $\boldsymbol{W_2}$ are no longer transposed
over adjacent frequency bands, since convolutions are merely performed over
the time variable.
The definition of $\boldsymbol{y_2}[t, k_1, k_2]$ rewrites as
\begin{IEEEeqnarray}{rCl}
\IEEEeqnarraymulticol{3}{l}{
\boldsymbol{y_2}[t,k_1,k_2]} \nonumber \\
& = & \boldsymbol{b_2}[k_2] +
\boldsymbol{W_2}[t, k_1, k_2] \overset{t}{\ast} \boldsymbol{x_1}[t, k_1]
\nonumber \\
& = &
\boldsymbol{b_2}[k_2] +
\sum_{\substack{0 \leq \tau < \Delta t}}
\! \! \! \! \!
\boldsymbol{W_2}[\tau, k_1, k_2]
\boldsymbol{x_1}[t-\tau, k_1],
\IEEEeqnarraynumspace
\label{eq:convolution1d}
\end{IEEEeqnarray}
and similarly for $\boldsymbol{y_3}[t, k_1, k_3]$.
While this approach is theoretically capable of encoding pitch invariants, it is
prone to early specialization of low-level features, thus
not fully taking advantage of the network depth.

However, the situation is improved if the feature maps
are restricted to the highest frequencies in the constant-Q spectrum.
It should be observed that, around the $n^{\textrm{th}}$ partial of a quasi-harmonic sound,
the distance in log-frequency between neighboring partials decays like $1/n$,
and the unevenness between those distances decays like $1/n^2$.
Consequently, at the topmost octaves of the constant-Q spectrum,
where $n$ is equal or greater than $Q$, the partials appear close to each other and almost
evenly spaced.
Furthermore, due to the logarithmic compression of loudness, the polynomial decay
of the spectral envelope is linearized: thus, at high frequencies, transposed pitches
have similar spectra up to some additive bias.
The combination of these two phenomena implies that the correlation between
constant-Q spectra of different pitches is greater towards high frequencies, and that
the learning of polyvalent feature maps becomes tractable.

In our experiments, the one-dimensional convolutions over the time variable
range from
$\mathrm{A_6}$ ($1.76\,\mathrm{kHz}$) to
$\mathrm{A_9}$ ($14\,\mathrm{kHz}$).

\subsection{Convolutions on the pitch spiral at low frequencies}
The weight sharing strategy presented above exploits the facts that,
at high frequencies, quasi-harmonic partials are numerous, and that the
amount of energy within a frequency band is independent of pitch.

At low frequencies, we claim
that the harmonic comb is sparse and covariant with respect to pitch shift.
Observe that, for any two distinct partials taken at random between $1$ and $n$,
the probability that they are in octave relation is slightly above $1/n$.
Thus, for $n$ relatively low, the structure of harmonic sounds is well
described by merely measuring correlations between partials one octave apart.
This idea consists in rolling up the log-frequency axis into a Shepard pitch spiral,
such that octave intervals correspond to full turns, hence aligning all coefficients
of the form $\boldsymbol{x_1}[t, k_1 + Q \times j_1]$ for $j_1 \in \mathbb{Z}$
onto the same radius of the spiral.
Therefore, correlations between power-of-two harmonics are revealed by
the octave variable $j_1$.

To implement a convolutional network on the pitch spiral, we crop the constant-Q
spectrogram in log-frequency into $J_1 = 3$ half-overlapping bands whose height
equals $2Q$, that is two octaves.
Each feature map in the first layer, indexed by $k_2$, results from the sum of  convolutions between a time-frequency kernel and a band, thus emulating a linear
combination in the pitch spiral with a 3-d tensor $\boldsymbol{W_2}[\tau, \kappa_1, j_1, k_2]$ at fixed $k_2$.
The definition of $\boldsymbol{y_2}[t, k_1, k_2]$ rewrites as
\begin{IEEEeqnarray}{rCl}
\boldsymbol{y_2}[t,k_1,k_2]
= & &
\! \! \! \! \! \! \! \! \! \! \! \! \! \! \! \! \! \! \! \!
\boldsymbol{b_2}[k_2]  \nonumber \\
& +
\! \sum_{\tau, \kappa_1, j_1} \! &
\boldsymbol{W_2}[\tau, \kappa_1, j_1, k_2] \nonumber \\
& &\times
\boldsymbol{x_1}[t - \tau, k_1 - \kappa_1 - Q j_1].
\IEEEeqnarraynumspace
\end{IEEEeqnarray}
The above is different from training two-dimensional kernel on
a time-chroma-octave tensor, since it does not suffer from artifacts
at octave boundaries.

The linear combinations of frequency bands that are one octave apart,
as proposed here,
bears a resemblance with engineered features for musical instrument
recognition \cite{Peeters2004}, such as tristimulus,
empirical inharmonicity, harmonic spectral deviation,
odd-to-even harmonic energy ratio, as well as
octave band signal intensities (OBSI) \cite{Joder2009}.

Guaranteeing the partial index $n$ to remain low is achieved by
restricting the pitch spiral to its lowest frequencies.
This operation also partially circumvents the problem of fixed spectral envelope
in musical sounds, thus improving the validness of the stationarity assumption.
In our experiments, the pitch spiral ranges from
$\mathrm{A_2}$ ($110\,\mathrm{Hz}$) to
$\mathrm{A_6}$ ($1.76\,\mathrm{kHz}$).

In summary, the classical two-dimensional convolutions make a stationarity assumption
among frequency neighborhoods. This approach gives a coarse approximation
of the spectral envelope.
Resorting to one-dimensional convolutions allows to disregard nonstationarity,
but does not yield a pitch-invariant representation per se:
thus, we only apply them at the topmost frequencies, \ie where the
invariance-to-stationarity ratio in the data is already favorable.
Conversely, two-dimensional convolutions on the pitch spiral addresses
the invariant representation of sparse, transposition-covariant spectra:
as such, they are best suited to the lowest frequencies,
\ie where partials are further apart and pitch changes can be approximated by log-frequency
translations.
The next section reports experiments on instrument recognition that capitalize
on these considerations.

\section{Applications}\label{sec:single-instrument}
The proposed algorithms are trained on a subset of MedleyDB v1.1. \cite{Bittner2014},
a dataset of 122 multitracks annotated with instrument activations.
We extracted the monophonic stems corresponding to a selection of eight pitched
instruments (see Table \ref{table:single-label-durations}).
Stems with leaking instruments in the background were discarded.

\begin{table}
    \begin{center}
    \begin{tabular}{|c|cc|cc|}
        \hline
        & minutes & tracks & minutes & tracks \\
        \hline
         piano & 58 & 28 & 44 & 15 \\
         violin & 51 & 14 & 49 & 22 \\
         dist. guitar & 15 & 14 & 17 & 11 \\
           female singer & 10 & 11 & 19 & 12 \\
        clarinet & 10 & 7 & 13 & 18 \\
        flute & 7 & 5 & 53 & 29 \\
        trumpet & 4 & 6 & 7 & 27 \\
        tenor sax. & 3 & 3 & 6 & 5 \\  
        \hline
        total & 158 & 88 & 208 & 139 \\
        \hline
    \end{tabular}
    \end{center}
    \caption{
    Quantity of data in the training set (left) and test set (right).
    The training set is derived from MedleyDB.
    The test set is derived from MedleyDB for distorted electric guitar and female singer,
    and from \cite{Joder2009} for other instruments.
    \label{table:single-label-durations}}
\end{table}

\begin{table*}[t]
    \begin{center}
    \setlength{\unitlength}{1cm}
    \begin{tabular}{|c|cccccccc|c|}
        \hline
        & piano & violin & dist.    & female & clarinet & flute & trumpet & tenor & average \\
        &           &          & guitar & singer   &           &          &              &  sax.  &                \\
        \hline
        bag-of-features
        & \textbf{99.7} & \textbf{76.2} & \textbf{92.7} & 81.6 & 49.9 & 22.5 & 63.7 & \hphantom{0}4.4 & 61.4 \\
        and random forest
        &\g{0.1}&\g{3.1}&\g{0.4}&\s{1.5}&\s{0.8}&\s{0.8}&\s{2.1}&\s{1.1} & \s{0.5}\\
        \hline
        spiral
        & 86.9 & 37.0 & 72.3 & 84.4 & 61.1 & 30.0 & 54.9 & 52.7 & 59.9 \\
        (36k parameters)
        &\s{5.8}&\s{5.6}&\s{6.2}&\s{6.1}&\s{8.7}&\s{4.0}&\s{6.6}&\s{16.4}&\s{2.4}\\
        \hline
        1-d
        & 73.3 & 43.9 & 91.8 & 82.9 & 28.8 & \textbf{51.3} & 63.3 & \textbf{59.0} & 61.8 \\
        (20k parameters)
        &\s{11.0}&\s{6.3}&\s{1.1}&\s{1.9}&\s{5.0}&\g{13.4}&\s{5.0}&\g{6.8}&\s{0.9} \\
        \hline
        2-d, $32$ kernels
        & 96.8 & 68.5 & 86.0 & 80.6 & 81.3 & 44.4 & 68.0 & 48.4 & 69.1 \\
        (93k parameters)
        &\s{1.4}&\s{9.3}&\s{2.7}&\s{1.7}&\s{4.1}&\s{4.4}&\s{6.2}&\s{5.3}&\s{2.0} \\
        \hline
        spiral \& 1-d
        & 96.5 & 47.6 & 90.2 & 84.5 & 79.6 & 41.8 & 59.8 & 53.0 & 69.1 \\
        (55k parameters)
        &\s{2.3}&\s{6.1}&\s{2.3}&\s{2.8}&\s{2.1}&\s{4.1}&\s{1.9}&\s{16.5}&\s{2.0}\\
        \hline
        spiral \& 2-d
        & 97.6 & 73.3 & 86.5 & \textbf{86.9} & \textbf{82.3} & 45.8 & 66.9 & 51.2 & 71.7\\
        (128k parameters)
        &\s{0.8}&\s{4.4}&\s{4.5}&\g{3.6}&\g{3.2}&\s{2.9}&\s{5.8}&\s{10.6}&\s{2.0}\\
        \hline
        1-d \& 2-d
        & 96.5 & 72.4 & 86.3 & 91.0 & 73.3 & 49.5 & 67.7 & 55.0 & 73.8 \\
        (111k parameters)
        &\s{0.9}&\s{5.9}&\s{5.2}&\s{5.5}&\s{6.4}&\s{6.9}&\s{2.5}&\s{11.5}&\s{2.3}\\
        \hline
        2-d \& 1-d \& spiral
        & 97.8 & 70.9 & 88.0 & 85.9 & 75.0 & 48.3 & 67.3 & \textbf{59.0} & \textbf{74.0} \\
        (147k parameters)
        &\s{0.6}&\s{6.1}&\s{3.7}&\s{3.8}&\s{4.3}&\s{6.6}&\s{4.4}&\g{7.3}&\g{0.6}\\
        \hline
        2-d, 48 kernels
        & 96.5 & 69.3 & 84.5 & 84.2 & 77.4 & 45.5 & \textbf{68.8} & 52.6 & 71.7\\
        (158k parameters)
        &\s{1.4}&\s{7.2}&\s{2.5}&\s{5.7}&\s{6.0}&\s{7.3}&\s{1.8}&\s{10.1}&\s{2.0}\\
        \hline
    \end{tabular}
    \end{center}
    \caption{Test set accuracies for all presented architectures. All convolutional layers
    have $32$ kernels unless stated otherwise.
    \label{table:results}}
\end{table*}

The evaluation set consists of 126 recordings of solo music collected by
Joder \etal \cite{Joder2009}, supplemented with
23 stems of electric guitar and female voice from MedleyDB.
In doing so, guitarists and vocalists were thoroughly put either in the training set or the test set,
to prevent any artist bias.
We discarded recordings with extended instrumental techniques, since they are
extremely rare in MedleyDB.
Constant-Q spectrograms from the evaluation set were split into half-overlapping,
3-second excerpts.

For the two-dimensional convolutional network, each of the two layers consists of
$32$ kernels of width $5$ and height $5$, followed by a max-pooling of width $5$
and height $3$. Expressed in physical units, the supports of the kernels are respectively equal to $116\,\mathrm{ms}$ and $580\,\mathrm{ms}$ in time, $5$ and $10$ semitones in frequency.
For the one-dimensional convolutional network, each of two layers consists of
$32$ kernels of width $3$, followed by a max-pooling of width $5$. Observe that the temporal
supports match those of the two-dimensional convolutional network.
For the convolutional network on the pitch spiral, the first layer consists of $32$ kernels
of width $5$, height $3$ semitones, and a radial length of $3$ octaves in the spiral.
The max-pooling operator and the second layer are the same as in the two-dimensional convolutional network.

In addition to the three architectures above, we build hybrid networks implementing
more than one of the weight sharing strategy presented above.
In all architectures, the densely connected layers have $K_4=64$ hidden units
and $K_5=8$ output units.

In order to compare the results against shallow classifiers, we also extracted a typical
"bag-of-features" over half-overlapping, 3-second excerpts in the training set.
These features consist of the means and standard
deviations of spectral shape descriptors, \ie centroid, bandwidth, skewness,
and rolloff; the mean and standard deviation of the zero-crossing rate in the time domain;
and the means of MFCC as well as their first and second derivative.
We trained a random forest of $100$ decision trees on the resulting feature vector
of dimension $70$, with balanced class probability.

Results are summarized in Table \ref{table:results}.
First of all, the bag-of-features approach presents large accuracy variations
between classes, due to the unbalance of available training data.
In contrast, most convolutional models, especially hybrid ones, show less correlation
between the amount of training data in the class and the accuracy.
This suggests that convolutional networks are able to learn polyvalent
mid-level features that can be re-used a test time to discriminate rarer classes.

Furthermore, 2-d convolutions outperform other non-hybrid
weight sharing strategies. However, a class with broadband
temporal modulations, namely the distorted electric guitar, is best classified
with 1-d convolutions.

Hybridizing 2-d with either 1-d or spiral convolutions provide consistent, albeit
small improvements with respect to 2-d alone. The best overall accuracy is reached
by the full hybridization of all three weight sharing strategies, because of a performance boost for the rarest classes.

The accuracy gain by combining multiple models could simply be the result of
a greater number of parameters. To refute this hypothesis, we train a 2-d convolutional
network with $48$ kernels instead of $32$, so as to match the budget of the full hybrid model, \ie about 150k parameters. The performance is certainly increased, but not up to the hybrid models involving 2-d convolutions, which have less parameters. Increasing the number of kernels even more cause the accuracy to level out and the
variance between trials to increase.

Running the same experiments with broader frequency ranges of 1-d and spiral convolutions often led to a degraded performance, and are thus not reported.

\section{Conclusions}
Understanding the influence of pitch in audio streams is paramount to the design of
an efficient system for automated classification, tagging, and similarity retrieval in music.
We have presented deep learning methods to address pitch invariance
while preserving good timbral discriminability.
It consists in training a feed-forward convolutional network over the constant-Q spectrogram,
with three different weight sharing strategies according to the type of input:
along time at high frequencies (above $2\,\mathrm{kHz}$),
on a Shepard pitch spiral at low frequencies (below $2\,\mathrm{kHz}$),
and in time-frequency over both high and low frequencies.

A possible improvement of the presented architecture would be to
place a third convolutional layer in the time domain before performing long-term
max-pooling, hence modelling the joint dynamics of the three mid-level feature maps.
Future work will investigate the association of the presented weight sharing strategies
with recent advances in deep learning for music informatics,
such as data augmentation \cite{McFee2015-muda},
multiscale representations \cite{Hamel2012, Anden2015},
and adversarial training \cite{Kereliuk2015}.

\bibliography{lostanlen_ismir2016}

\end{document}